# Development of Compact Muon Spectrometer Using Multiple Pressurized Gas Cherenkov Radiators

Junghyun Bae[1,*] and Stylianos Chatzidakis[1]
[1] School of Nuclear Engineering, Purdue University, IN 47906
bae43@purdue.edu, schatzid@purdue.edu

**Abstract**

In both particle physics and muon applications, a high-resolution muon momentum measurement capability plays a significant role not only in providing valuable information on the properties of subatomic particles but also in improving the utilizability of cosmic ray muons. Typically, muon momentum is measured by reconstructing a curved muon path using a strong magnetic field and muon trackers. Alternatively, a time-of-flight and Cherenkov ring imager are less frequently applied, especially when there is a need to avoid a magnetic field. However, measurement resolution is much less than that of magnetic spectrometers, approximately 20% whereas it is nearly 4% or less when using magnets and trackers. Here, we propose a different paradigm to estimate muon momentum that utilizes multiple pressurized gas Cherenkov radiators. Using the fact that the refractive index of gas medium varies depending on its pressure and temperature, we can optimize the muon Cherenkov threshold momentum levels for which a muon signal will be detected. In this work, we demonstrate that muon momentum can be estimated with mean resolution of $\sigma_P/p < 20\%$ and mean classification rate of 90.08% in the momentum range of 0.1 to 10.0 GeV/c by analyzing optical photon signals from each Cherenkov radiator. We anticipate our new spectrometer will significantly improve quality of imaging and reduce scanning time in cosmic ray muon applications by being incorporated with existing instruments.

[*]Corresponding author
Email: bae43@purdue.edu

## 1. Introduction

Muons are one of the most versatile particles in both particle physics and nuclear engineering. Not only are muons the key particles in accelerator and neutrino studies, e.g., CMS and miniBooNE [1,2], but also a promising non-conventional radiographic probe in muon applications, e.g., nuclear reactor and waste imaging [3–7], homeland security [8–14], geotomography [15–19], and archeology [20,21]. In particle physics, two important quantities, muon trajectory and momentum, need to be measured. Typically, muon trackers and a combination of strong magnets are used to reconstruct muon trajectories and measure muon momentum, respectively [22]. In certain cases where application of a strong magnetic field is not practical, a time-of-flight or Cherenkov ring imager is used to measure muon momentum [23]. However, the average measurement resolution is much lower than that of magnetic spectrometers, ~20% or worse [24].

When it comes to muon tomography, or muography, the imaging resolution is often limited by the naturally low muon flux, approximately $7.25 \pm 0.1$ cm$^{-2}$s$^{-1}$sr$^{-1}$ at sea level [25]. In addition, muon flux varies with zenith angle and detector configuration [26,27]. Incorporating additional information such as muon momentum has potential to maximize and expand utilizability of cosmic ray muons to various engineering applications. Despite the apparent benefits [28,29], it is still very challenging to measure muon momentum in the field without deploying large and expensive spectrometers. Existing spectrometers are not well suited for muon tomographic applications that demand portability, limited available space, small size requirements, and minimal cost. Additional challenges with existing spectrometers include interference with muon traveling path (mostly a problem for magnetic spectrometers), a vast array of detectors, e.g., Ring Image Cherenkov Detector, or a very long line of sight (for time-of-flight spectrometers). At present, neither fieldable nor portable muon spectrometers exist.

In this work, we present a different paradigm for muon momentum measurement using multiple pressurized gas Cherenkov radiators. By carefully selecting gas pressure at each radiator, we can optimize muon Cherenkov momentum threshold for which a muon signal will be detected. In this design, a muon passing through all radiators will activate only those that have a threshold momentum level less than actual muon momentum. By measuring the presence of Cherenkov radiation signals in each radiator, we can estimate the muon momentum with absolute resolution of ±0.5 GeV/c using six Cherenkov radiators over the muon momentum range of 0.1 to 10.0 GeV/c. In addition, the dimension of our proposed spectrometer is smaller than 1 m which renders it easily transportable for field measurements. The primary benefits of our concept are: (i) it offers a versatile and portable muon spectrometer that provides similar or better resolution with existing spectrometers, (ii) it extends the measurable momentum range without relying on bulky and expensive magnetic spectrometers, and (iii) it provides a solution to muon momentum measurement in the field that is needed for various engineering applications, e.g., archaeology, geotomography, cosmic radiation measurements in the International Space Station, off-site particle experiment laboratories, nuclear safeguards, and non-destructive monitoring techniques.

## 2. Design and concept of Cherenkov muon spectrometer

To develop a compact Cherenkov muon spectrometer, we use multiple pressurized gas Cherenkov radiators with different refractive indices. We carefully select the refractive index of each radiator by varying the gas pressure to achieve the necessary Cherenkov threshold muon momentum. Muon momentum measurement resolution is determined by the maximum threshold momentum level and the number of radiators. A fine resolution can be obtained by increasing the number of radiators in a given measurable muon momentum range. The absolute momentum measurement resolution, $\sigma_p$, is written by:

$$\Delta p_{th} = \frac{p_{th,max}}{N_{rad} - 1} \quad (1)$$

$$\sigma_p = \pm \Delta p_{th}/2 \quad (2)$$

where $N_{rad}$ is the number of radiators. $\Delta p_{th}$ and $p_{th,max}$ are the interval and maximum threshold momentum, respectively. When the actual muon momentum is $p_\mu$, muon momentum can be estimated using:

$$\hat{p}_\mu = (N_i - 1)\Delta p_{th} + \sigma_p \quad (3)$$

where $N_i$ is the largest radiator number that emits Cherenkov radiation when $i = 1, 2, \ldots, N_{rad}$. For our prototype design, we used six radiators, comprising of a glass ($SiO_2$) and five pressurized gas radiators ($CO_2$) with linearly increasing threshold momentum levels, $p_{th} = 0.1, 1.0, 2.0, 3.0, 4.0$, and $5.0$ GeV/c. The maximum momentum level is limited to 10.0 GeV/c because approximately 84% of cosmic ray muons have momentum less than 10.0 GeV/c as shown in Fig. 1. $CO_2$ is chosen as a gas radiator because it allows to be pressurized up to nearly 5.7 MPa and can cover a wide range of momentum threshold levels. The lowest momentum threshold radiator is replaced by $SiO_2$ because the threshold of 0.1 GeV/c is not possible to achieve using high-pressure $CO_2$ gas. Each radiator is designed to be triggered when a muon has at least momentum higher than a pre-selected level in that particular radiator. In other words, depending on the muon momentum, none to all Cherenkov radiators can be triggered. An overview of measuring muon momentum of $\sigma_p = \pm 0.5$ GeV/c using six Cherenkov radiators (one $SiO_2$ and five $CO_2$ gases) is illustrated in Fig. 2.

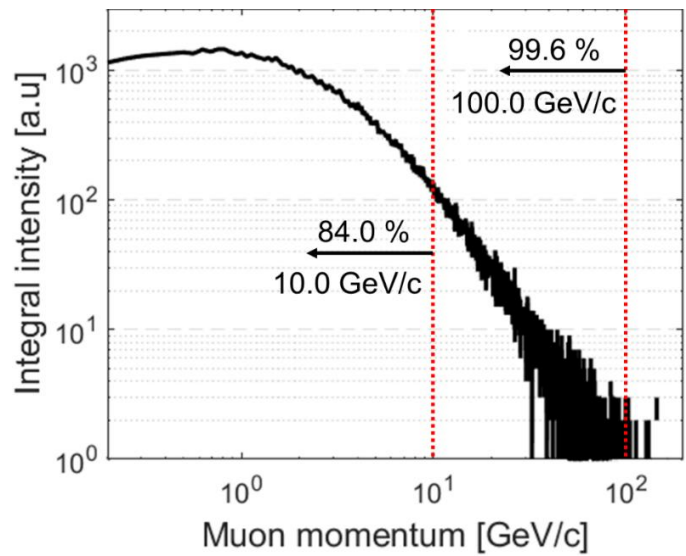

**Fig. 1.** Approximate fraction of muons within the cosmic ray muon spectrum.

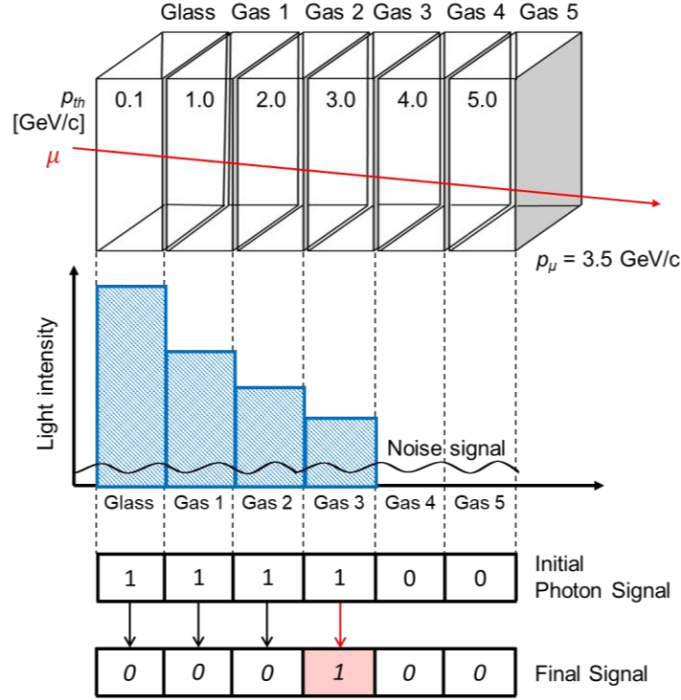

**Fig. 2.** Overview of measuring muon momentum with a resolution, $\sigma_p = \pm 0.5$ GeV/c using six Cherenkov radiators, one solid ($SiO_2$) and five gas ($CO_2$) radiators. Light intensity decreases as threshold momentum level increases because it depends on the refractive index of radiator. When Cherenkov radiation signals are detected in the radiator, it records "1" otherwise "0". Subtracting the recorded signals allows identification of the radiator with the highest momentum threshold and classification of the muon momentum within a predetermined momentum range. For example, if a muon with actual momentum 3.5 GeV/c passes through the radiators, the muon spectrometer will assign this muon in the muon momentum range $3.0 < p_\mu$ (GeV/c) $< 4.0$ estimating its momentum as 3.5 GeV/c [30].

*2.1. Selection of gas radiators*

To utilize gas Cherenkov radiators in our proposed Cherenkov spectrometer, we rely on two theories in physics: (i) Cherenkov effect and (ii) Lorentz-Lorenz equation [31,32]. When the velocity of a muon exceeds the speed of light in an optically transparent medium, or a radiator, Cherenkov radiation is emitted. This happens when the following criterion is satisfied:

$$\beta_\mu n > 1, \qquad (4)$$

where $\beta_\mu$ is the ratio of the velocity of muon in the radiator to that of light in a vacuum ($= v_\mu/c$) and $n$ is the refractive index of the radiator. As shown in Eq. (4), a minimum muon velocity is required so that Cherenkov light is emitted in a given radiator. A threshold momentum for muons can be derived as [33]:

$$p_{th} c = \frac{m_\mu c^2}{\sqrt{n^2 - 1}}, \qquad (5)$$

where $p_{th}$ is the threshold muon momentum and $m_\mu c^2$ represents the muon rest mass energy ($\cong 105.66$ MeV). From Eq. (5), we can see that the muon threshold momentum can be adjusted by properly selecting the refractive index of radiator. For a gas medium, the refractive index of gas varies depending on both pressure and temperature. When $n^2 \approx 1$, the refractive index of gas radiator can be approximated by Lorentz-Lorenz equation:

$$n \approx \sqrt{1 + \frac{3 A_m p}{RT}}, \qquad (6)$$

where $A_m$ is the molar refractivity, $p$ is the gas pressure, $T$ is the absolute temperature, and $R$ is the universal gas constant. By substituting Eq. (6) into Eq. (5), Cherenkov threshold momentum for muons in terms of both gas pressure and temperature can be written as:

$$p_{th} c = m_\mu c^2 \sqrt{\frac{R}{3 A_m} \frac{T}{p}} \qquad (7)$$

From Eq. (7), it can be seen that a threshold momentum is proportional to $p^{-1/2}$. As a result, by changing gas pressure, the threshold momentum can be changed without the need to replace any materials. The variation of Cherenkov threshold muon momentum, $p_{th}$, and the refractive index, $n$, as a function of gas pressure for four radiators, $C_3F_8$, $C_3H_2F_4$ (R1234yf), $C_4F_{10}$, and $CO_2$ is shown in Fig. 3. We vary the gas pressure with maintaining at room temperature because (i) changing pressure covers a wider $p_{th}$ range over that of temperature and (ii) maintaining high-pressure is more practical than keeping very low gas temperature in the field. Properties of gas Cherenkov radiators are summarized in Table I. $C_4F_{10}$ and $CO_2$ have been used as gas Cherenkov radiators in the Jefferson Lab [34]. A refrigerant, R1234yf, is a promising substitute radiator to replace R12 ($CCl_2F_2$) due to it lower Ozone Depletion Potential (ODP) [35]. $C_3F_8$ is alternative to $C_4F_{10}$ because $C_4F_{10}$ cannot be pressurized higher than 380 kPa without condensation at room temperature. Based on the results from Fig. 3 and Table I, we chose $CO_2$ gas as our Cherenkov gas radiator because it covers a wide range of threshold momentum and it is commercially available in a large quantity at minimal cost. $C_3F_8$ would be also be a good candidate but it has a smaller vapor pressure and is more expensive.

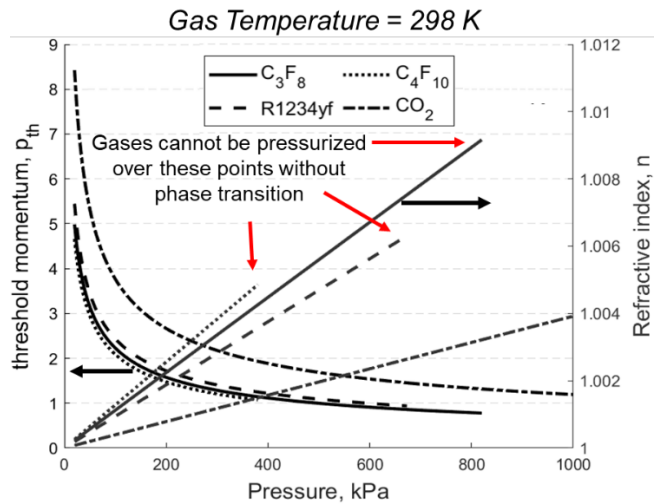

**Fig. 3.** Cherenkov threshold muon momentum (left) and refractive index (right) for $C_3F_8$, R1234yf, $C_4F_{10}$, and $CO_2$ gas radiators as a function of gas pressure. Note: gases cannot be pressurized above their vapor pressure without condensation.

**Table 1**

Material properties for selected Cherenkov gas radiators at room temperature [35–37].

| Selected gas radiators | $C_3F_8$ | R1234yf | $C_4F_{10}$ | $CO_2$ |
|---|---|---|---|---|
| Vapor pressure [MPa] | 0.820 | 0.673 | 0.380 | 5.7 |
| Vapor density [kg/m$^3$] | 12.5 | 37.6 | 24.6 | 1.977 |
| Molecular weight [g/mol] | 188.02 | 114.04 | 236.03 | 44.01 |
| Refractive index [−] | 1.0011 | 1.0010 | 1.0015 | 1.00045 |
| Polarizability, $\alpha$ [× $10^{-30}$ m$^3$] | 7.4 | 6.2 | 8.44 | 2.59 |

## 2.2. Cherenkov radiation

Typically, there are three predominant photon emission mechanisms when muons interact with optically transparent medium: (i) Cherenkov radiation, (ii) scintillation, and (iii) transition radiation [38]. Cherenkov photon intensity depends on various factors, refractive index, particle velocity, and path length. Expression for Cherenkov photon intensity was first derived by I. Frank and I. Tamm [39]. According to the Frank-Tamm formula, the expected number of Cherenkov photons in unit path length, $dN_{ch}/dx$, between wavelength, $\lambda_1$ and $\lambda_2$, within a spectral region is given by:

$$\frac{dN_{ch}}{dx} = 2\pi\alpha \int_{\lambda_1}^{\lambda_2} \left(1 - \frac{1}{n^2(\lambda)\beta^2}\right)\frac{d\lambda}{\lambda^2} \tag{8}$$

where $\alpha$ is the fine structure constant ($\cong 1/137$), $\beta$ is the muon phase velocity, and $n(\lambda)$ is the refractive index of gas radiator. When electromagnetic wavelength range is limited to optical photons (200–700 nm), Eq. (8) is simplified by:

$$\frac{dN_{ch}}{dx} \cong 1150 \sin^2 \theta_c \tag{9}$$

$$\theta_c = \cos^{-1}\left(\frac{1}{\beta n}\right), \tag{10}$$

where $\theta_c$ is the Cherenkov angle.

Cherenkov photon emission is the result of instant physical disorder caused by a charged particle such as a muon. Estimated duration of Cherenkov light flash, $\Delta t$ is given by [40,41]:

$$\Delta t = \frac{r}{\beta c}\sqrt{\beta^2 n(\lambda) - 1}\bigg|_{\lambda_1}^{\lambda_2}, \qquad (11)$$

where *r* is the distance to the observer. Calculated Cherenkov light flash duration for a high-energy muon is less than a nanosecond as shown in Fig. 4.

*2.3. Scintillation*

When a portion of muon energy is transferred to the medium, molecules and atoms are excited against one another. The excitation energy is then released as a form of light which is called scintillation, when they return to the ground state. Wavelength of scintillation photon is extended approximately between 200 and 700 nm depending on types of scintillation medium [42]. Light yield of scintillation depends on muon energy deposited in the radiators. The mean number of scintillation photons per unit distance, *dN$_{sc}$/dx*, is given by [43]:

$$\frac{dN_{sc}}{dx} = S\frac{dE/dx}{1 + k_B(dE/dx)}, \qquad (12)$$

where *S* is the scintillation efficiency, $k_B$ is the Birks' coefficient, and *dE/dx* is the muon mass stopping power which can be calculated using the Bethe equation [44]. When the scintillation medium is gaseous state and E ≥ 300 keV, then $k_B \approx 0$.

Unlike Cherenkov radiation, light flash duration of scintillation is associated with photon absorption, excitation, and relaxation. The time response of the prompt fluorescence of scintillation is given by [45]:

$$I/I_0 = f(t)e^{-t/\tau} \qquad (13)$$

where *I/I$_0$* is the normalized light intensity, *f(t)* represents the characteristic Gaussian function, and *τ* is the time constant describing decay. The scintillation light flash duration is mainly determined by a decay constant which is order of *μsec* for inorganic and few *nsec* for organic scintillation materials [42]. Estimated flash time responses and electromagnetic wavelength spectra for both Cherenkov and scintillation are shown in Fig. 4. Scintillation photons can be systematically discriminated from Cherenkov photons using two methods: differences in (i) light flash timing and (ii) light emission spectrum. A significant timing difference is observed between scintillation and Cherenkov radiation and they can be separated using a fast and fine timing resolution [46] as shown in Fig. 4 (left). In addition, given that scintillation and Cherenkov radiation are emitted in a homogeneous material, their spectra are distinguishable and most scintillation photons can be filtered out by selectively choosing either a UV photodiode detector or IR photovoltaic detector as shown in Fig. 4 (right).

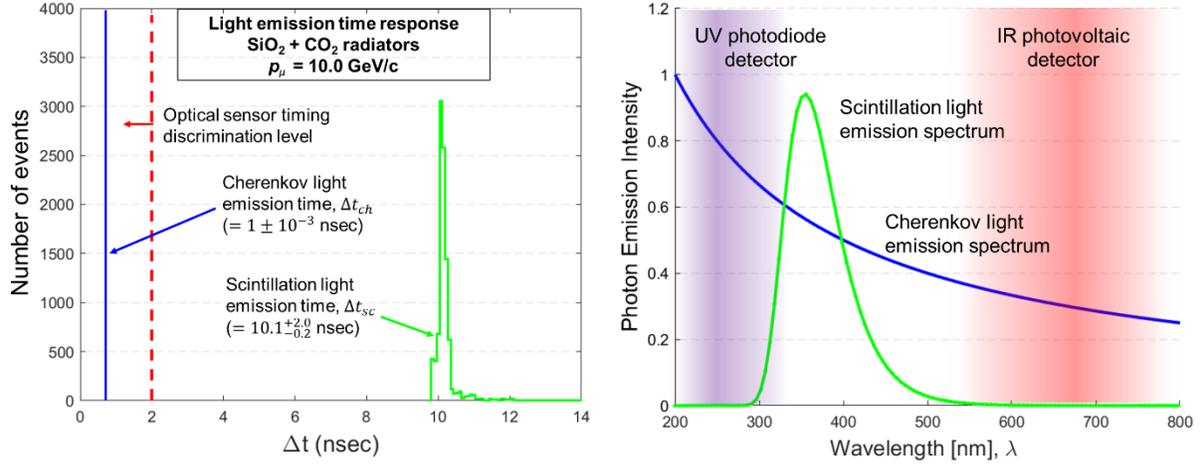

**Fig. 4.** Light emission time responses from Geant4 simulations for Cherenkov radiation and scintillation in $SiO_2$ and $CO_2$ radiators (left) and typical spectra of Cherenkov radiation and scintillation (right) [47]. Examples of sensitive EM wavelength for both UV photodiode and IR photovoltaic detectors to selectively detect Cherenkov radiation are also presented (right).

*2.4. Transition radiation*

When a muon passes through inhomogeneous media such as a boundary between two radiators, photons are emitted and they are called transition radiation. Transition radiation is sometimes misinterpreted as Cherenkov radiation because both have directional photon emission as shown in Fig. 5. However, transition radiation is neither related to particle energy loss by collisions at boundaries nor deceleration of muons. Especially, transition radiation that emits photons with a visible wavelength is called optical transition radiation (OTR) and a mean yield of OTR, when $\gamma \gg 1$ is given by [48]:

$$N_{tr} = \frac{z^2 \alpha}{\pi}\left[(\ln \gamma - 1)^2 + \frac{\pi^2}{12}\right], \quad (14)$$

where $z$ is the muon charge, $\alpha$ is the fine structure constant, and $\gamma$ is the Lorentz factor. Expected optical photon intensity is not significant ($10^{-3} \sim 10^{-4}$) when the number of physical boundaries is small and momentum range is not greater than 10 GeV/c [49].

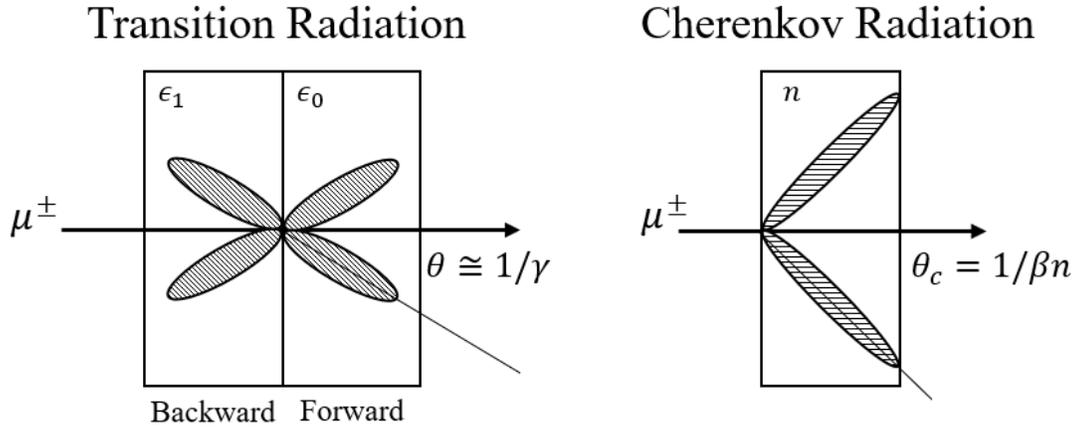

**Fig. 5.** Characteristics of transition and Cherenkov radiation when a muon passes through the media (photon emission angles are exaggerated) [50].

*2.5. Optical photon emission in a gas radiator*

Within each radiator, Cherenkov radiation, scintillation, and transition radiation can be emitted depending on conditions of muons and radiators. When incident muon momentum exceeds a threshold momentum, Cherenkov photons are expected to be emitted along a muon traveling path with a characteristic angle, or Cherenkov angle, $\theta_c$. On the other hand, scintillation and transition photons can be emitted regardless of incoming muon momentum. Characteristics of photon emissions by Cherenkov radiation, scintillation, and transition radiation in both gas radiator A ($p_\mu > p_{th}$) and B ($p_\mu < p_{th}$) are shown in Fig. 6.

Expected optical photon yields by Cherenkov, scintillation, and transition radiation and Signal-to-Noise Ratio (SNR), $N_{ch}/(N_{sc}+N_{tr})$, as a function of radiator length for pressurized $CO_2$ and $C_3F_8$ are shown in Fig. 7. It appears that Cherenkov light emission yield increases more rapidly than that of scintillation. This means that to achieve a higher SNR, a large gas radiator would be preferable. In addition, the expected number of optical photons by transition radiation is negligible.

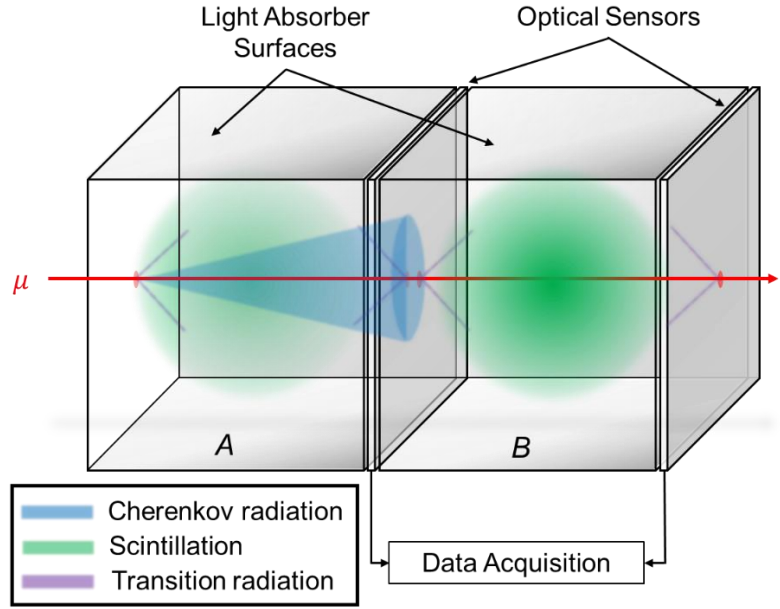

**Fig. 6.** Characteristics of light emission by Cherenkov radiation, scintillation, and transition radiation in two radiators. Radiator A (left) emits Cherenkov photons since $p_\mu > p_{th}$ whereas radiator B (right) does not because $p_\mu < p_{th}$. All surfaces of radiator containers are covered by light absorber except one in which optical sensors are placed. It is noted that the Cherenkov and transition radiation have characteristic directional light emission whereas scintillation light uniformly emits in all directions.

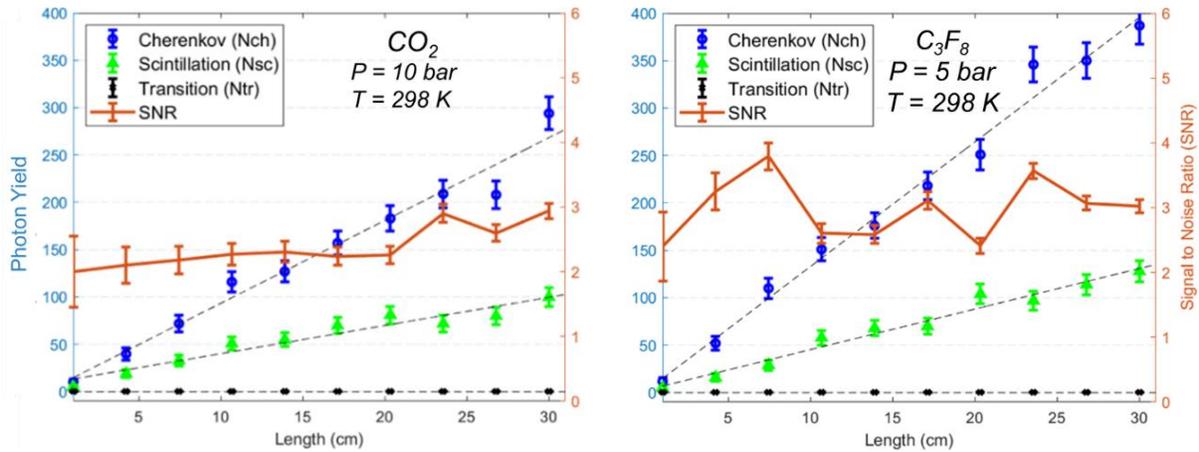

**Fig. 7.** Expected photon emission intensities by Cherenkov radiation, scintillation, and transition radiation when $p_\mu$ = 4.0 GeV/c as a function of length in two pressurized gas radiators, $CO_2$ (left) and $C_3F_8$ (right). They show that Cherenkov light emission yield increases rapidly than those of scintillation and transition radiation. Error bars represent 1σ.

### 3. Geant4 Modeling

*3.1. Geometry and materials*

Geant4 (Geometry ANd Tracking) is a computational tool developed by CERN for Monte-Carlo particle transport simulations in many applications such as high-energy, nuclear, accelerator, medical, and space physics [51,52]. We used Geant4 to simulate muon interactions with matter, secondary particle production, scattering, absorption, and light emission in our proposed Cherenkov muon spectrometer. A surface area of radiators is $20 \times 20$ cm$^2$ and gas radiators are enclosed by an aluminum housing with a thickness of 1 cm. Length of a glass radiator (SiO$_2$) is 1 cm, gas radiators are 10 cm each, and the aluminum photon absorber is 1 mm. All radiators and absorbers are included in the "world" volume which is made of air. Five pressurized CO$_2$ gas radiators are placed next to a solid radiator side by side. Each radiator chamber includes pressurized CO$_2$ gas to provide a pre-selected refractive index. Some CO$_2$ gas radiators are depressurized to achieve $p_{th}$ = 4 and 5 GeV/c because Cherenkov threshold momentum for muon at atmospheric pressure is 3.52 GeV/c. Main parameters used in simulations for solid and gas radiators are summarized in Table II. Aluminum is added to simulate a strong photon absorber. Once optical photons arrive at the absorber surface, all disappear.

**Table 2**

Properties and parameters of materials used in Geant4 simulations.

| Radiator ID | 1 | 2 | 3 | 4 | 5 | 6 | Absorber |
|---|---|---|---|---|---|---|---|
| Material | SiO$_2$ | CO$_2$ | CO$_2$ | CO$_2$ | CO$_2$ | CO$_2$ | Al |
| $<Z/A>$ [-] | 0.4973 | 0.4999 | 0.4999 | 0.4999 | 0.4999 | 0.4999 | 0.4818 |
| Length [cm] | 1 | 10 | 10 | 10 | 10 | 10 | 0.1 |
| $p_{th}$ [GeV/c] | 0.1 | 1.0 | 2.0 | 3.0 | 4.0 | 5.0 | - |
| Refractive index [-] | 1.45 | 1.00557 | 1.00139 | 1.00062 | 1.00035 | 1.00022 | - |
| Pressure [bar] | - | 14.4857 | 3.6214 | 1.6095 | 0.9054 | 0.5794 | - |
| Density [kg/m$^3$] | 2500 | 27.83 | 6.55 | 2.88 | 1.61 | 1.03 | 2700 |
| Radiation Length [cm] | 10.69 | 1.93E4 | 1.96E4 | 1.96E4 | 1.97E4 | 1.97E4 | 8.90 |

*3.2. Cosmic ray muons and physics list*

All muons were initially generated 10 cm away from the center of solid radiator surface. All major physics, i.e., three optical photon emission mechanisms by both primary muons and secondary particles, scattering and absorption, decays, energy loss, are included in Geant4 reference physics list, QGSP_BERT [38]. All optical photons are accordingly tagged by types of mother particle (primary muon or secondary particle) and light emission mechanisms (Cherenkov, scintillation, and transition radiation).

### 4. Analytical model benchmarking

To verify our Geant4 model, we performed two benchmarking simulations: cosmic ray muon (i) scattering displacement distribution and (ii) energy loss. Analytical models for these benchmarking examples were successfully developed based on multiple Coulomb scattering (MCS) and Bethe equation, respectively.

*4.1. Scattering angle distribution*

When a muon travels through matter, it is randomly deflected due to Coulomb interactions with atomic nuclei and electrons. The result of MCS is approximated using Gaussian distribution and its root mean square (rms) is derived by Highland [53].

$$f(\theta|0, \sigma_\theta^2) = \frac{1}{\sqrt{2\pi}\sigma_\theta} e^{-\theta^2/2\sigma_\theta^2} \qquad (15)$$

$$\sigma_\theta = \frac{13.6 \text{ MeV}}{\beta c p} \sqrt{\frac{X}{X_0}} \left[1 + 0.088 \log_{10}\left(\frac{X}{X_0}\right)\right], \qquad (16)$$

where $\theta$ and $\sigma_\theta$ are cosmic ray muon scattering angle and rms of scattering angle, respectively. $X_0$ is the radiation length and $X$ is the length of the scattering medium. When a muon travels in inhomogeneous materials, the effective radiation length is given by:

$$\frac{X_{total}\rho_e}{X_{0,e}} = \sum_i \frac{X_i \rho_i}{X_{0,i}} \qquad (17)$$

where $X_{total}$ is the total length, $\rho_e$ is the effective density of materials. $X_i$, $X_{0,i}$ and, $\rho_i$ are the length, radiation length, and density of $i^{th}$ material component. All parameters used to compute effective radiation length, $X_{0,e}$ are summarized in Table II. In addition, the rms of plane displacement of muons, $\sigma_{plane}$ is related to $\sigma_\theta$:

$$\sigma_{plane} = X\sigma_\theta \qquad (18)$$

The analytical calculation of muon displacement is shown as a circle centered at the origin (initial muon x-y coordinate). The estimated radii (1σ, 2σ, and 3σ) of muon scattering displacement distributions for selected muon energies, 0.5, 1.0, 4.0, and 10.0 GeV, and Geant4 simulations are in good agreement and the results are shown in Fig. 8.

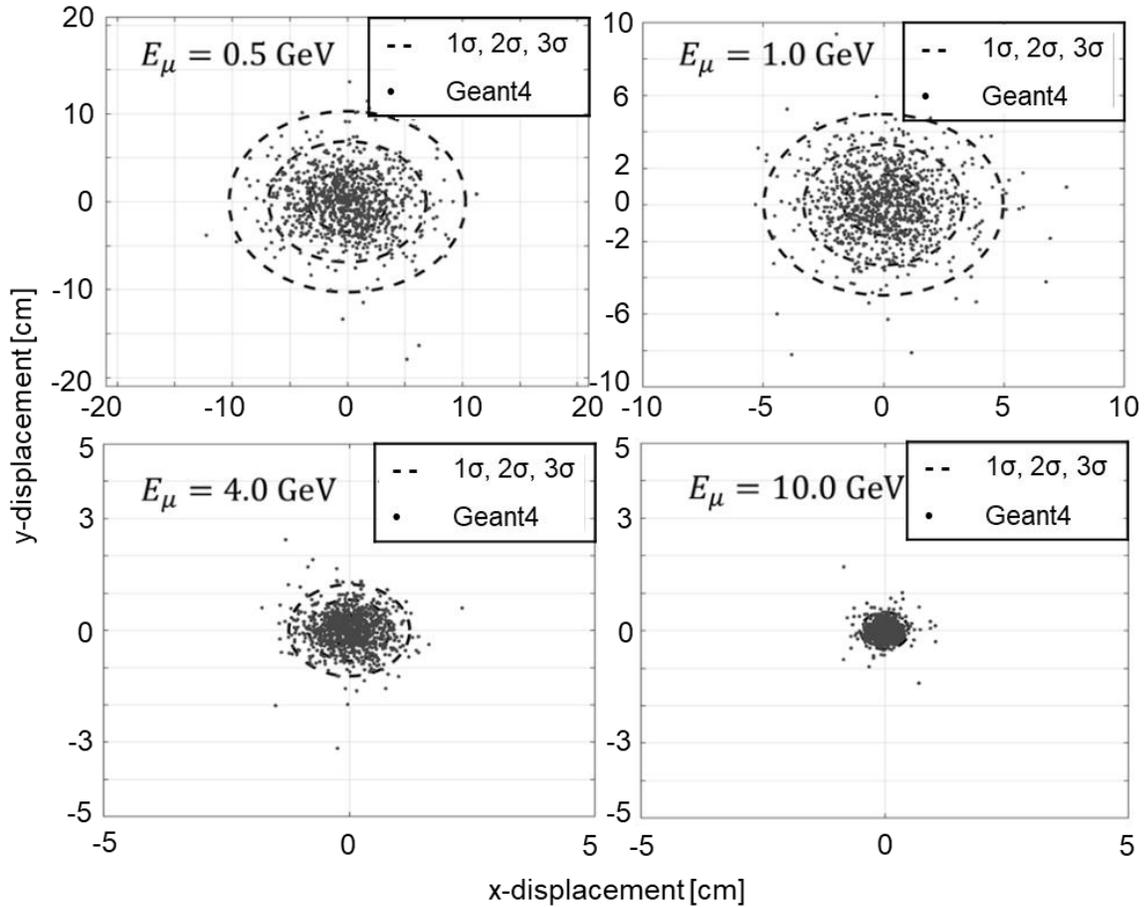

**Fig. 8.** The results of muon scattering displacement distributions in Geant4 simulations and analytical estimation using MCS Gaussian approximation. Each projected Gaussian circle (2D Gaussian distribution) represents 1σ (inner), 2σ (middle), and 3σ (outer). 99% of simulated muons are enclosed in the 3σ Gaussian circle. It is noted that different range of x- and y-axes are used.

*4.2. Energy loss*

Since muon energy loss by ionization dominates in most\r cosmic ray muon energy range (0.1 to 100 GeV/c), a mean muon mass stopping power can be estimated by using Bethe equation [54]. Fig. 9 shows Geant4 simulation results of muon energy loss in radiators as a function of muon energy using $10^4$ muon samples. Even though the amount of muon energy loss varies slightly from 7 to 12 MeV depending on muon energy, a fraction of energy loss to initial muon energy is insignificant, < 1%. The results from Geant4 simulations and analytical estimations using Bethe equation are in good agreement with each other.

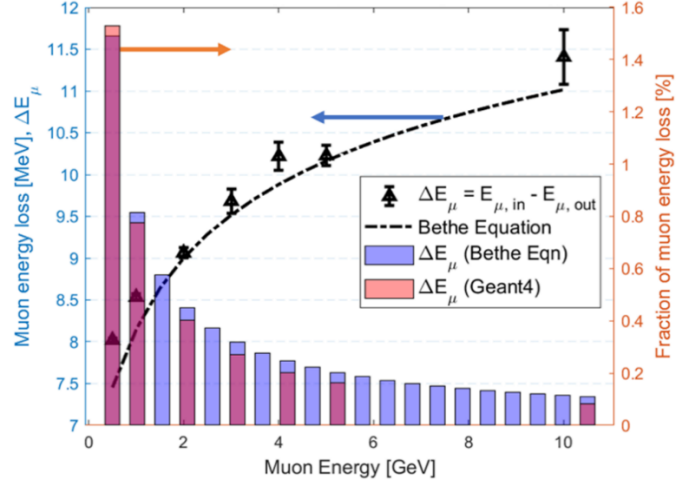

**Fig. 9.** Estimated cosmic ray muon energy loss using Bethe equation and Geant4 simulations using $10^4$ muon samples (left) and fraction of the energy loss to incident energy of cosmic ray muons as a function of initial muon energy using Geant4 and Bethe equation (right).

## 5. Results

### 5.1. Geant4 simulations

Various mono-energetic muons are generated to evaluate a functionality of our proposed Cherenkov muon spectrometer in Geant4 simulations. Measured photon signals are classified into two categories, Cherenkov photons (signals) and others (noise). In Geant4 simulations, three mono-energetic muons, $p_\mu$ = 0.5, 3.1 and 10.0 GeV/c, horizontally enter Cherenkov muon spectrometer and travel through all radiators as shown in Fig. 10. When $p_\mu$ = 0.5 GeV/c, only the first radiator emits Cherenkov radiation because muon momentum is greater than the threshold momentum of first radiator. Similarly, the first four and all radiators emit conical-shaped Cherenkov radiation when $p_\mu$ = 3.1 and 10.0 GeV/c, respectively. In addition to Cherenkov light, scintillation photons are observed in some radiators as shown in Fig. 10 even when $p_\mu < p_{th}$. Scintillation photons can be efficiently discriminated from Cherenkov radiation using (i) characteristic photon emission direction ($\theta c$ vs $4\pi$) and (ii) light flash timing, and (iii) photon emission spectrum. Furthermore, a scintillation photon yield is not significant in all radiators except solid radiator. No transition radiation is observed in Fig. 10 because it is a highly rare event for a few GeV muons and for the small number of boundary system.

It is noteworthy that a negatively charged particle production (a trajectory in red except a muon) in the second radiator is observed in Fig. 10-(c). Optical photons occur due to either Compton scattering or Cherenkov photon emissions by secondary particles, mainly electrons, which can be produced by radiative muon decays and muon to electron conversions [55]:

$$\mu^- \rightarrow e^- \bar{\nu}_e \nu_\mu \tag{19}$$

$$\mu N_{Al} \rightarrow e N_{Al} \tag{20}$$

Eq. (19) represents a radiative decay of $\mu^-$ with a mean lifetime of $\tau_\mu \cong 2.2$ μsec. Radiative muon decays are a primary source of secondary charged particles, electrons. Eq. (20) represents a muon to electron conversion (mu2e). Muons can be captured by the *1s* orbital of aluminum, then mono-energetic electron emits with an absence of neutrino [56].

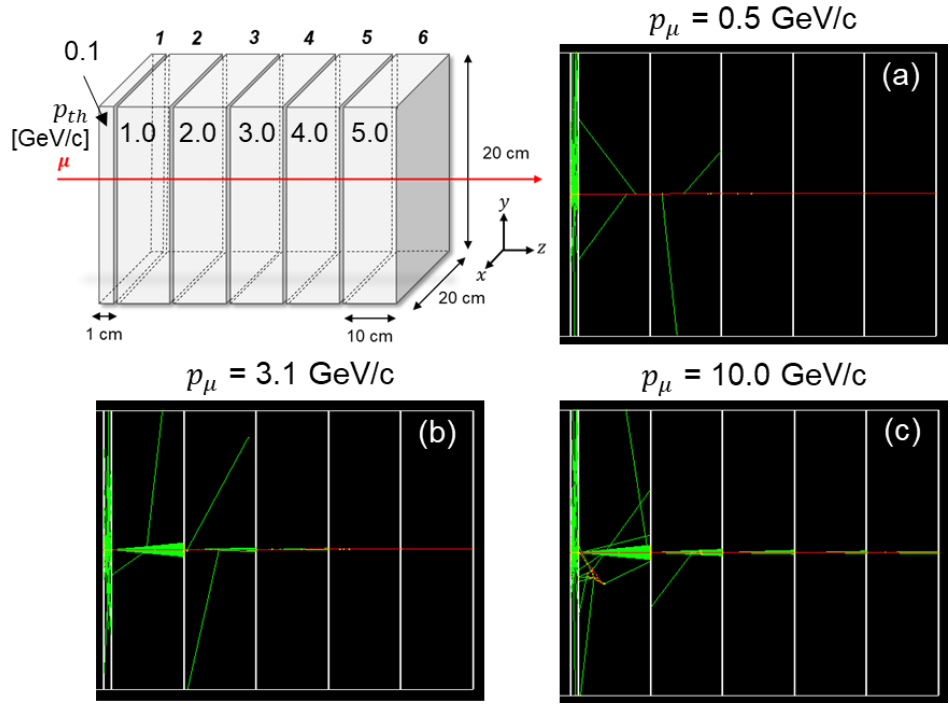

**Fig. 10.** Overview of Cherenkov muon spectrometer and visualized Geant4 simulations when a single negatively charged muon horizontally enters in the center. Three mono-energy muons are simulated: (a) $p_\mu$ = 0.5 GeV/c, (b) $p_\mu$ = 3.1 GeV/c, and (c) $p_\mu$ = 10.0 GeV/c. Cherenkov radiation is observed only when muon momentum exceeds a threshold momentum level of radiators. Unlike Cherenkov radiation, scintillation photons are emitted in all directions. Secondary particle ($e^-$) is also observed in (c) due to either a muon radiative decay or muon to electron conversion. Note: red and green represent negatively charged particles and photons, respectively.

*5.2. Performance evaluation*

We evaluated the performance of our proposed Cherenkov muon spectrometer by reconstructing a sea-level cosmic ray muon spectrum. Analytical models and experimental results of a cosmic ray muon spectrum at sea level can be found in literature [57–59]. We used an open-source Monte Carlo muon generator ("Muon_generator_v3" [60]) which is developed based on semi-empirical model by Smith and Duller [61]. The result of reconstructed cosmic ray muon spectrum using six momentum groups is shown in Fig. 11. The last bin in histogram shows a disagreement with the actual cosmic ray muon spectrum because all muons that have momentum greater than 5.0 GeV/c are categorized in ">5.0 GeV/c" bin. This maximum measurable momentum level problem can be resolved by extending the threshold momentum range and increasing the number of radiators. However, although the increased number of radiators within a limited overall size improves momentum measurement resolution, it will negatively impact the SNR and degrade the measurement quality (accuracy) due to decreased Cherenkov photon signals in each radiator. In other words, the optimal number of radiators is a balance between two tradeoffs and it must be selected in accordance with conditions such as required size, resolution, and accuracy.

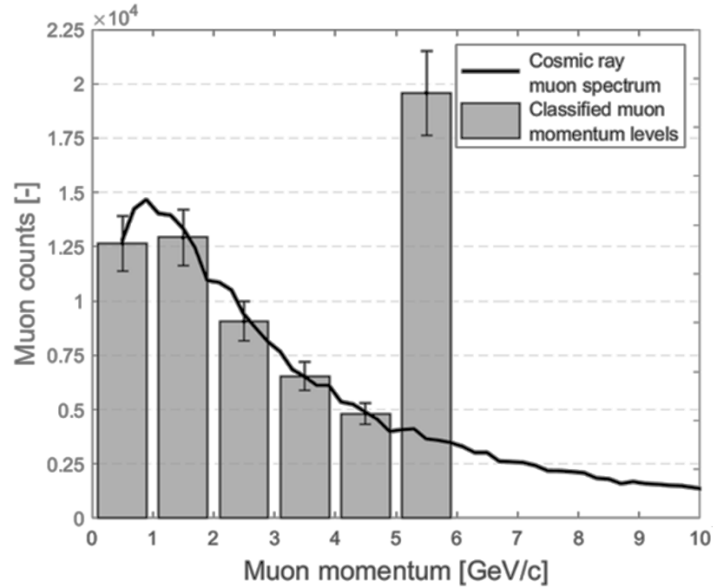

**Fig. 11.** Reconstructed cosmic ray muon spectrum using the proposed Cherenkov spectrometer with six radiators. Error bars represents 1σ.

*5.3. Classification rate*

To evaluate the accuracy of our proposed Cherenkov muon spectrometer, a concept of classification rate (CR) is described. It represents the probability that a momentum estimator from signal readout correctly indicates the actual muon momentum range:

$$CR = \frac{\text{True Positive Classification}}{\text{Positive Classification}} \qquad (21)$$

where Positive Classification represents the recorded muon counts in a certain momentum class and True Positive Classification represents the expected muon counts when actual incoming muon momenta are given. Geant4 simulation results for optical photon signal responses from all radiators when $p_\mu = 1.1$ and 3.1 GeV/c ($N_\mu = 10^4$) are shown in Fig. 12. A significant difference in photon counts between when $p_\mu > p_{th}$ and $p_\mu < p_{th}$ is observed and it can be utilized to estimate actual muon momentum by analyzing photon signals from each radiator. However, due to a presence of noise, the expected number of optical photons for all radiators (SiO$_2$ and five pressurized CO$_2$) are non-zero even when $p_\mu = 1.1$ GeV/c. Because noise signals in undesirable radiators, e.g., scintillation and Cherenkov radiation by secondary particles, are much smaller than primary Cherenkov radiation signals, we can filter them out by deducting a certain level of counts from total photon signals, a technique also known as signal discriminator. The use of both signal discriminator levels of 1 and 2 to eliminate noise signal when $p_\mu$ = 1.1 and 3.1 GeV/c is shown in Fig. 12. In both examples, noise signals are efficiently removed by a level-2 discriminator.

Further, an analysis of the classification rate for muon momentum range from 0.1 to 10.0 GeV/c is performed using $10^4$ mono-energetic muon samples in Geant4 simulations. The results are shown in Fig. 13. In the low muon momentum range (< 1.0 GeV/c), CR is 70–80% with a level-2 discriminator due to strong scintillation photon signals in the glass radiator. In the intermediate momentum range ($1.0 < p_{th} <$ 7.0 GeV/c), the discriminator plays a significant role to maintain the CR level greater than 90%. In the high momentum range ($7.0 < p_{th} < 10.0$ GeV/c), however, the CR level decreases because the Cherenkov photon intensity is too low in rarefied $CO_2$ gas radiators (approximately < 0.5 atm). Therefore, using a discriminator must be avoided in this range. When a combination of various discriminator levels is applied, i.e., a level-2 discriminator for low and intermediate momentum ranges, level 0 or 1 for a high momentum range, then we can achieve a mean classification rate of 90.08 %. The classification rate can be further increased using additional signal processing techniques, e.g., timing discrimination.

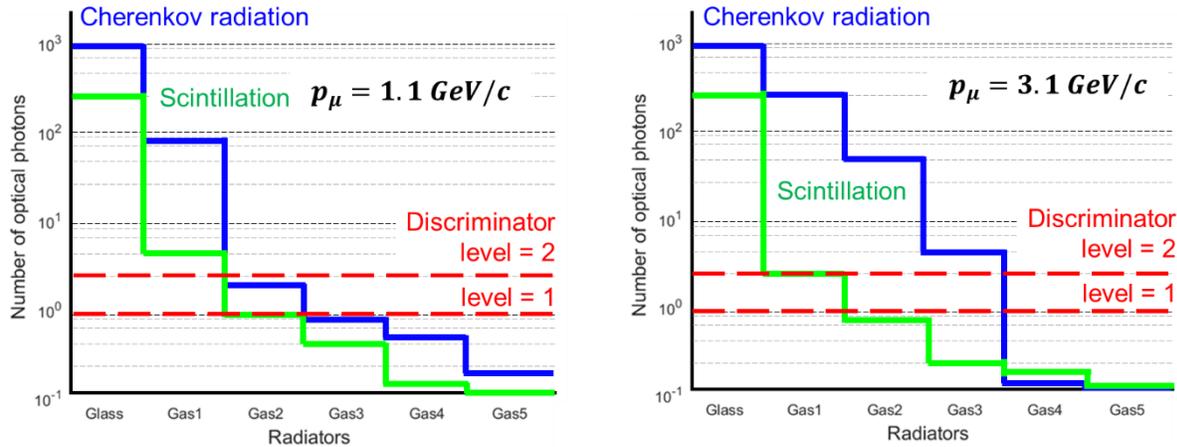

**Fig. 12.** Geant4 simulation results for optical photon signals from all radiators ($SiO_2$ and five pressurized $CO_2$) when $p_\mu$ = 1.1 (left) and 3.1 GeV/c (right) ($N_\mu = 10^4$). A noticeable difference in photon counts between when $p_\mu > p_{th}$ and $p_\mu < p_{th}$ is observed. The use of both signal discriminator levels of 1 and 2 to eliminate noise signal is also presented.

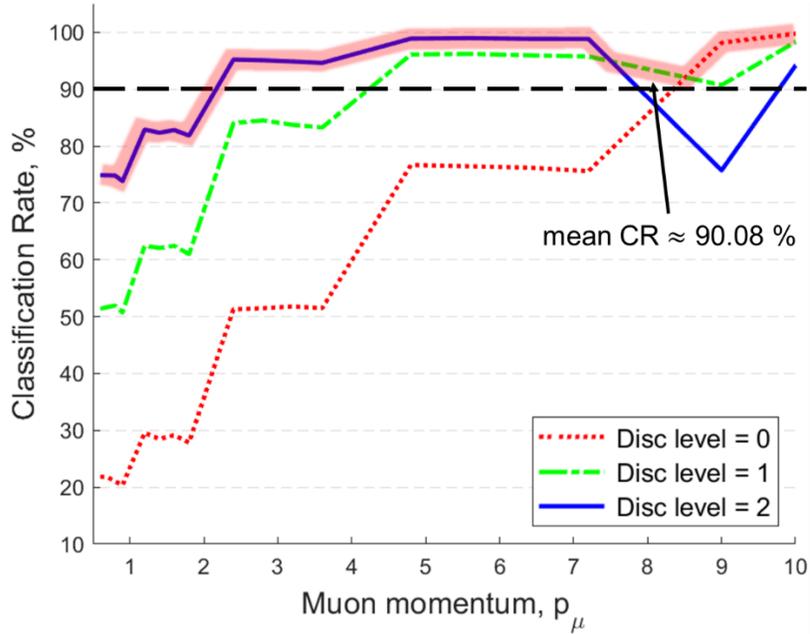

**Fig. 13.** Gean4 simulation results for classification rates of proposed Cherenkov muon spectrometer as a function of muon momentum from 0.1 to 10.0 GeV/c using $10^4$ muons. Various signal discriminator levels are used to minimize noise signals. A mean CR using a combination of discriminator level of 0, 1, and 2 is 90.08 % (highlighted plot).

### 6. Conclusion

In this paper, we presented a new concept for muon momentum measurement using multiple pressurized gas Cherenkov radiators and performed a detailed feasibility study using analytical models and Geant4 Monte-Carlo muon transport simulations. To eliminate the need for magnetic or time-of-flight spectrometers, our muon spectrometer relies on pressurized gas Cherenkov radiators. In this work, six sequential threshold momentum levels were selected, 0.1, 1.0, 2.0, 3.0, 4.0, and 5.0 GeV/c. When a muon passes all radiators, none to all Cherenkov radiators can emit Cherenkov radiation depending on the actual muon momentum. By monitoring photon signal readout from optical sensors in each radiator, we can find out radiators that emit Cherenkov radiation. Then, we can estimate the actual muon momentum by analyzing digital signals from radiators with a designed momentum resolution.

Our results demonstrate that muon momentum can be estimated with a resolution of ±0.5 GeV/c using 6 radiators, one $SiO_2$ and five pressurized $CO_2$, over the range of 0.1 to 5.0 GeV/c. Despite the presence of various noise sources, e.g., scintillation, transition radiation, and secondary particles (mainly electrons), we successfully classified the actual muon momentum by analyzing recorded photon signals. To improve measurement accuracy, a signal discriminator is devised to efficiently eliminate noise from signals. The performance of our proposed muon spectrometer was evaluated using classification rate over muon momentum range from 0.1 to 10.0 GeV/c. The results showed a mean classification rate of approximately 90.08 %. The performance of our Cherenkov muon spectrometer can be further improved by applying additional techniques to discriminate scintillation from Cherenkov radiation, e.g., timing discrimination. Finally, the momentum measurement resolution and the momentum range can be improved by increasing the number of radiators.

**CRediT authorship contribution statement**

**J. Bae**: Resource, Methodology, Supervision, Numerical and analytical computation, Writing – original draft. **S. Chatzidakis**: Investigation, Supervision, Project administration, Funding acquisition, Writing – review & editing.

**Declaration of competing interest**

The authors declare that they have no known competing financial interests or personal relationships that could have appeared to influence the work reported in this paper.

**Data availability**

The data that support the findings of this study are available from the corresponding author upon reasonable request.

**Acknowledgments**

This research is being performed using funding from the Purdue Research Foundation.